\documentclass[aps,preprint,amssymb,12pt,floatfix]{revtex4}
\usepackage{subfigure}
\usepackage{lscape,graphicx}
\usepackage{rotating}

\begin{document}
\title{Dynamics of allosteric transitions in GroEL}
\author{Changbong Hyeon$^1$, George H. Lorimer$^{1,2}$ and D. Thirumalai$^{1,2}$}
\thanks{Corresponding author phone: 301-405-4803; fax: 301-314-9404; thirum@glue.umd.edu}
\affiliation{$^1$Biophysics Program\\
Institute for Physical Science and Technology\\
$^2$Department of Chemistry and Biochemistry\\
University of Maryland, College Park, MD 20742\\
}

\date{\today}
\baselineskip = 22pt

\begin{abstract}
The chaperonin GroEL-GroES, a machine which helps some proteins to fold, cycles through a number of allosteric states, the $T$ state, with high affinity for substrate proteins (SPs), the ATP-bound $R$ state, and the $R^{\prime\prime}$ ($GroEL-ADP-GroES$) complex.  
Structures are known for each of these states. 
Here, we use a self-organized polymer (SOP) model for the GroEL allosteric states and a general structure-based technique to simulate the dynamics of allosteric transitions in  two subunits of GroEL and the heptamer.  
The $T \rightarrow R$ transition, in which the apical domains undergo counter-clockwise motion, is mediated by a multiple salt-bridge switch mechanism, in which  a series of salt-bridges break and form.  
The initial event in the  $R \rightarrow R^{\prime\prime}$ transition, during which GroEL rotates clockwise, involves a spectacular outside-in movement of  helices K and L that results in  K80-D359 salt-bridge formation.  
In both the transitions there is considerable heterogeneity in the transition pathways. 
The transition state ensembles (TSEs) connecting the $T$, $R$, and $R^{\prime\prime}$ states are broad with the the TSE for the $T \rightarrow R$ transition being more plastic than the $R\rightarrow R^{\prime\prime}$ TSE.  
The results suggest that GroEL functions as a force-transmitting device in which forces of about 
(5-30) pN may act on the SP during the reaction cycle.
\end{abstract}
\maketitle

\newpage

{\bf INTRODUCTION}\\

The hallmark of allostery in biomolecules is the conformational changes at distances far from the sites at which ligands bind \cite{Perutz89QuartRevBiophys,Changeux05Science}. 
Allosteric transitions arise from  long-range spatial correlations between a network of residues.  
 Signaling between the residues likely results in conformational transitions. 
The potential link between large scale allosteric transitions and function is most vividly illustrated in biological nanomachines \cite{Horovitz05COSB}.  For example, during transcription DNA polymerases undergo a transition from open to closed state that is triggered by dNTP binding \cite{Doublie99Structure}.  
Although allosteric transitions were originally associated with multi-subunit assemblies, it is now believed that a network of residues encode the dynamics of monomeric globular proteins \cite{Kern03COSB}.  

Computational methods have been used to determine the network of spatially correlated residues  (wiring diagram) that trigger the functionally relevant conformational changes.  
Recently, static methods,  either sequence \cite{Lockless99Science,Kass02Proteins,Dima06ProtSci} or structure-based \cite{Zheng05Structure,Zheng06PNAS}  have been proposed to predict the allosteric wiring diagram.  
However, in order to fully understand the role of allostery it is important to dynamically monitor the structural changes that occur in the transition from one state to another \cite{Miyashita03PNAS,Best05Structure,Bahar05COSB,Okazaki06PNAS,Koga06PNAS}.   
Here, we propose a method for determining the allosteric mechanism in biological systems 
with applications to dynamics of such processes 
in the chaperonin GroEL, an ATP-fueled nanomachine, 
which facilitates folding of proteins (SPs) that are otherwise destined to aggregate \cite{ThirumARBBS01,Horwich06ChemRev}. 

GroEL  consists of two heptameric rings,  stacked back-to-back.  
Substrate proteins are captured by  GroEL in the \textit{T} state (Fig. 1) while  
 ATP-binding  triggers a transition to the \textit{R} state.
The $T$ and $R$ structures show that the equilibrium $T \Leftrightarrow R$ transition results 
in large (nearly rigid body) movements of the apical (A) domain (Fig. 1), and somewhat smaller changes in the conformations of the intermediate (I) domain. Binding of the co-chaperonin GroES requires dramatic movements in the A domains which  double the volume of the central cavity.  Comparison of the structures of the $T, $ $R$, and the $R^{\prime\prime}$  ($GroEL-(ADP)_7-GroES$)  indicates that the equatorial (E) domain, which serves as an anchor \cite{ThirumARBBS01}, undergoes comparatively fewer structural changes.
Although structural and mutational studies \cite{HorovitzJMB94,Horovitz01JSB,Yifrach95Biochem} have  identified many important residues that affect GroEL function, only few studies have explored the dynamics of allosteric transitions between the various states \cite{Ma98PNAs,KarplusJMB00,BaharMSB06}.  

Here, we use the self-organized polymer (SOP) model of GroEL and a novel technique (Methods) 
to monitor the order of events in the  $T \rightarrow R$, $R \rightarrow R^{\prime\prime}$, and 
$T \rightarrow R^{\prime\prime}$ transitions.  By simulating the dynamics of ligand-induced conformational changes in the heptamer and two subunits we have obtained an unprecedented
view of the key interactions that drive the various allosteric transitions. The dynamics of transition between the states are achieved (Methods) under the assumption that the rate of conformational changes is slower than the rate  at which ligand-binding induced strain propagates. 
Because of the simplicity of the SOP model we have been able to generate multiple trajectories to resolve the key  events in the allosteric transitions.  We make a number of predictions including the identification of a multiple salt-bridge switch mechanism in the $T \rightarrow R$ transition,  and the occurrence of dramatic movement of helices K and L in the $R \rightarrow R^{\prime\prime}$ transition.
The structures of the  transition state ensembles  that connect the various end points show considerable variability mostly localized in the A domain.\\  

{\bf RESULTS and DISCUSSION}\\

{\bf Heptamer dynamics show that the A domains rotate counter-clockwise in the $T \rightarrow R$ transition and clockwise in the $R \rightarrow R^{\prime\prime}$ transition:}
In order to probe the global motions in the various stages of GroEL allostery we simulated the entire heptamer (Methods).  
The dynamics of the $T \rightarrow R$ transition, monitored using the time-dependent changes in the angles $\alpha$, $\beta$, and $\gamma$ (see the caption in Fig. 2 for definitions), that measure the relative orientations of the subunits,  show (Fig. 2-A) that the A domains twist in a counter-clockwise manner in agreement with experiment \cite{Ranson01Cell}. The net changes in the angles in the $R \rightarrow R^{\prime\prime}$ transition, which occurs in a clockwise direction (Fig. 2-B), is greater than in the $T \rightarrow R$ transition. As a result the global  $T \rightarrow R^{\prime\prime}$ transition results in a net $\approx 110^o$ rotation of the A domains. Surprisingly, there are large variations in the range of angles explored by the individual subunits during the $T \rightarrow R \rightarrow R^{\prime\prime}$ transitions.  There are many more inter-subunit contacts in the E domain than in the A domain, thus permitting each A domain to move more independently of one another. Fig. 2 shows that the dynamics of each subunit is distinct despite the inference, from the end states alone, 
that the overall motion occurs without significant change in the root mean square deviation (RMSD) of the individual domains. 
The time-dependent
changes in the angles $\alpha$, $\beta$, and $\gamma$  from one subunit to another are indicative of
an inherent dynamic asymmetry in the individual subunits that has been noted in static structures \cite{Danzinger03PNAS,Shimamura04Structure}. 
As in the $T \rightarrow R$ transition,  there is considerable dispersion in the time-dependent changes in  $\alpha$, $\beta$, and $\gamma$ 
of the individual subunits  (Fig. 2-B) during the $R\rightarrow R^{\prime\prime}$ transition. 

The clockwise rotation of apical domain alters the nature of lining of the SP binding sites (domain color-coded in magenta in Fig. 1). 
The dynamic changes in the $\gamma$ angle  (Fig. 2) associated with the hinge motion of the I domain that is perpendicular to the A domain lead to an expansion of  the 
overall volume of the heptamer ring. 
More significant conformational changes, that lead to doubling of the volume of the cavity,  take place in the $R\rightarrow R^{\prime\prime}$ transition. 
The apical domain is erected, so that the SP binding sites are oriented upwards providing binding 
interfaces for GroES. Some residues, notably  357-361, which are completely exposed on the exterior surface in the $T$ state move to the interior surface during the $T \rightarrow R \rightarrow R^{\prime\prime}$ transitions.\\

{\bf Global $T \rightarrow R$ and $R \rightarrow R^{\prime \prime}$ transitions follow two-state kinetics:}  
Time-dependent changes in RMSD with respect to a reference state ($T$, $R$, or $R''$), from which a specific allosteric transition commences  (Fig. 3), 
differ from molecule-to-molecule, which reflects the heterogeneity of the underlying dynamics (Fig. 3-A). 
Examination  of the RMSD for a particular trajectory in the transition region (Fig. 3-A inset) shows that the molecule undergoes multiple passages through the transition state (TS).  Assuming that RMSD is a reasonable reaction coordinate, we find that GroEL spends a substantial fraction of time (measured with respect to the first passage time) in the TS region during the $T \rightarrow R$ transition.  
By averaging over 50 individual trajectories we find that the ensemble average of the time-dependence of RMSD for both the  $T\rightarrow R$ and 
$R\rightarrow R''$ transitions follow single exponential kinetics.  From this observation we conclude that, despite a broad transition region, the allosteric transitions can be approximately described by a two-state model.  
From the exponential fits of the global RMSD changes  we find that the relaxation times for the two transitions are  $\tau_{T\rightarrow R}\sim 25 \mu s$ and $\tau_{R\rightarrow R^{\prime\prime}}\sim 71\mu s$. 
The larger time constant for the $R\rightarrow R^{\prime\prime}$ transition compared to the $T\rightarrow R$ transition is due to the substantially greater rearrangement of the structure in the former.
Unlike the global dynamics characterizing the overall  motion of GroEL, 
the local dynamics describing the formation and rupture of key interactions that drive GroEL allostery cannot be described using two-state kinetics (see below). \\ 

{\bf $T\rightarrow R$ transition is triggered by downward tilt of helices F and M in the I-domain followed by a multiple salt-bridge switching mechanism: }
Several residues in helices  F (141-151) and  M (386-409) in the I domain interact  with the nucleotide-binding sites in the E domain thus creating a tight nucleotide binding pocket.  The favorable interactions are enabled by the F, M helices tilting by about $15^o$ that results in the closing of the nucleotide-binding sites.  A number of residues around the nucleotide binding pocket are highly conserved \cite{Brocchieri00ProtSci,Stan03BioPhysChem}.
Since the $T \rightarrow R$ transition involves the formation and breakage of intra- and inter- subunit contacts we simulated two interacting subunits  so as to dissect the order of events.

(i) The ATP-binding-induced downward tilt  of the  F, M helices is the earliest event \cite{KarplusJMB00}  that accompanies the subsequent spectacular movement of GroEL.  The changes in the angles F and M helices make
with respect to their orientations in the $T$ state occur  in concert  (Fig. 3-C). 
At the end of the $R \rightarrow R^{\prime \prime}$ transition the helices have tilted on average by about $25^o$ in all  (Fig. 3-C). 
There are variations in the extent of the tilt depending on the molecule (see inset in Fig. 3-C). 
Upon the downward tilt of the F and M helices, the entrance to the ATP binding pocket  narrows as evidenced by the rapid decrease in the distance  
between P33  and N153 (Fig. 4).  A conserved residue, P33  contacts  ATP in the $T$ state and ADP in the $R^{\prime\prime}$ structure and is involved in allostery \cite{BaharMSB06}.   
The contact number of N153 increases substantially as a result of loss in accessible surface area during the $R \rightarrow R^{\prime \prime}$ 
transition \cite{Stan03BioPhysChem}. 
In the $T$ state E386, located at the tip of M helix, forms inter-subunit salt-bridges with R284, R285, and R197. 
In the transition to the $R$ state these salt-bridges are  disrupted  and the  formation of a new intra-subunit salt-bridge with K80 takes place simultaneously (see the middle panel in Fig. 4). 
The dynamics show that the tilting of M helix must precede the formation of inter-subunit salt-bridge between the charged residues E386 with K80.

(ii) The rupture of the intra-subunit salt-bridge  D83-K327 occurs nearly simultaneously with the
disruption of the E386-R197 inter-subunit interaction.  The 
distance between the $C_{\alpha}$ atoms of D83 and K327 of around  8.5 \AA\  in the $T$ state 
slowly increases to the equilibrium distance of $\sim 13$ \AA\ in the R state with relaxation time $\tau\approx 100$ $\mu s$  (the top panel in Fig. 4). The establishment of K80-E386 salt-bridge occurs around the same time as the rupture of R197-E386 interaction.
 In the $T \rightarrow R$ formation a network of salt-bridges are broken and new ones formed (see below). 
At the residue level, the reversible formation and breaking of D83-K327 salt-bridge, in concert with
the inter-subunit salt-bridge switch associated with E386 \cite{Ranson01Cell} and E257 \cite{Stan05JMB,Danzinger06ProtSci}, are among the most significant  events that dominate the $T \rightarrow R$ transition. 

Remarkably, the coordinated global motion is orchestrated by a multiple salt-bridge switching mechanism.  
The movement of the  A domain results in the dispersion  of the SP binding sites  (Fig. 1) and  also  leads to the rupture of the  E257-R268 inter-subunit salt-bridge. The kinetics of breakage of the E257-R268 salt-bridge are distinctly non-exponential (the last panel in Fig. 4).  
It is very likely that the dislocated SP binding sites maintain their stability through the inter-subunit salt-bridge formation between the apical domain residues.  To maintain the stable configuration in the R state, E257 engages in salt-bridge formation with positively charged residues that are initially buried at the interface
of inter-apical domain in the T state.  Three positively  charged residue at the interface of the apical domain in the R state, namely, 
K245, K321, and R322 are the potential candidates for the salt-bridge with E257.  During the $T \rightarrow R$ transitions E257 interact partially with K245, K321, and R322 as evidenced by the decrease in their distances (the last panel in the middle column of Fig. 4). The distance between E409-R501 salt-bridge, involving residues  that connect and hold the  I and E domains, remains intact at a distance  $\sim 10$ \AA\ throughout the whole allosteric transitions. 
This salt-bridge and two others (E408-K498 and E409-K498) might be important for enhancing positive intra-ring cooperativity and for stability of the chaperonins.  Indeed, mutations at sites E409 and R501 alter the stability of the various allosteric states \cite{Aharoni97PNAS}.  In summary, we find a coordinated dynamic changes in the network of salt-bridges are linked in the $T\rightarrow R$ transition.  

The order of events, described above, are not followed in all the trajectories.   
Each molecule follows somewhat different pathway during the allosteric transitions which is indicated by the considerable dispersion in the dynamics. However, when the time traces are averaged over a large enough sample, the global kinetics can be analyzed using a two state model.    
\\

{\bf  $R\rightarrow R^{\prime\prime}$ transition involves a spectacular outside-in movement of K and L helices accompanied by inter-domain salt-bridge formation K80-D359: }
The dynamics of the irreversible $R\rightarrow R''$  transition is propelled by substantial movements in the A domain helices K and L  
 that drive the dramatic conformational change in GroEL resulting in doubling of the volume of the cavity. 
The dynamics of the  $R \rightarrow R^{\prime\prime}$ transition also occur in stages.

(i) Upon ATP hydrolysis the F, M helices rapidly tilt by an additional $10^o$ (Fig. 3-C). Nearly simultaneously  there is a small  reduction in P33-N153 distance (7 \AA $\rightarrow$ 5 \AA) (see top panel in Fig. 5). These relatively small changes are the initial events in the $R \rightarrow R^{\prime\prime}$ transition. 

(ii) In the next step, the A domain undergoes significant conformational changes that are most vividly captured by the outside-in concerted movement of helices K and L. Helices K and L, that tilt by about $30^o$ during the $T \rightarrow R$ transition, further rotate by an additional $40^o$ when the $R \rightarrow R^{\prime\prime}$ transition occurs (Fig. 3-D). In the process, a number of largely polar and charged residues that are exposed to the exterior in the $T$ state line the inside of the cavity in the $R^{\prime\prime}$ state. The outside-in motion of K and L helices leads to an inter-domain salt-bridge K80-D359 whose $C_{\alpha}$ distance changes rapidly   
from about 40 \AA\ in the $R$ state to about 14 \AA\ in the $R^{\prime \prime}$  (Fig. 5). 

The wing of the apical domain that protrudes outside the GroEL ring in the $R$ state moves inside the cylinder. 
The outside-in motion facilitates the K80-D359 salt-bridge formation which in turn orients the position of the wing.  
The  orientation of the apical domain's wing inside the cylinder exerts a substantial strain (data not shown) on the 
GroEL structure. 
To relieve the strain,  the apical domain is forced to undergo  
a dramatic 90$^o$ clockwise rotation and 40$^o$ upward movement with respect to the $R$ state. 
As a result, the SP binding sites (H, I helices) are oriented in 
the upward direction.
Before the strain-induced alterations are possible the 
distance between K80 and D359 decreases drastically from that in R state (middle panel in Fig. 5). 
The clockwise motion of the apical domain occurs only after the formation of salt-bridge between K80 and D359.  On the time scale during which K80-D359 salt-bridge forms,  
the rupture kinetics of several inter-apical domain salt-bridges involving residues K245, E257, R268, K321, and R322, follow complex kinetics 
(Fig. 5). 
Formation of contact between I305 and A260 (a binding site for substrate proteins), and inter-subunit residue pair located at the interface of two adjacent apical domains in the  $R^{\prime\prime}$ state, occurs extremely slowly  compared to others.  
The non-monotonic and lag-phase kinetics observed in the rupture and formation of a number of contacts suggests that 
intermediate states must exist in the pathways connecting the $R$ and $R^{\prime \prime}$ states . 

 The clockwise rotation of apical domain, that is triggered by a network of 
salt-bridges as well as interactions between hydrophobic residues at the interface of subunits,  orients it in   the upward direction so as to permit the binding of the mobile loop of GroES. 
Hydrophobic interactions between SP binding sites and GroES drive the  $R\rightarrow R^{\prime\prime}$ transition. 
The hydrophilic residues, that are hidden on the side of apical domain in the $T$ or the $R$ state, are now exposed to form an interior surface of the GroEL (see the residue colored in yellow on the A domain in Fig. 1).  
The E409-R501 salt-bridge formed between I and A domains close to the $\gamma-P_i$ binding site 
is maintained  throughout the allosteric transitions including in the transition state  \cite{Aharoni97PNAS}.
 \\

{\bf Transition state ensembles (TSEs) are broad: }
The structures of the TSEs 
connecting the $T$, $R$, and $R^{\prime \prime}$ states are obtained using RMSD as a surrogate reaction coordinate.  
We assume that, for a particular trajectory,  the TS location is reached when  $\delta^{\ddagger} = |(RMSD/T)(t_{TS})-(RMSD/R)(t_{TS})|<r_c$
where $r_c = 0.2$ \AA, and $t_{TS}$ is the time at which  $\delta^{\ddagger} < r_c$. 
Letting the value of RMSD at the TS 
be $\Delta^{\ddagger} = 1/2\times|(RMSD/T)(t_{TS})+(RMSD/R)(t_{TS})|$ the
distributions $P(\Delta^{\ddagger})$ for $T \rightarrow R$ and $R \rightarrow R^{\prime\prime}$ transitions are broad (see Fig. S3 in the Supplementary Information). 
If $\Delta^{\ddagger}$ is normalized by the RMSD between the two end point structures to produce a Tanford $\beta$-like parameter $q^{\ddagger}$ (see caption to Fig. 6 for definition), 
we find that the width of the TSE for the $R\rightarrow R^{\prime\prime}$ is less than the $T \rightarrow R$ transition (Fig. 6-A). 
The mean values of $q^{\ddagger}$ for the two transitions show that the most probable TS is located close to the $R$ states in both $T \rightarrow R$ and $R \rightarrow R^{\prime\prime}$ transitions.

Disorder in the TSE structures (Fig. 6)  is largely localized in the 
A domain which shows that the substructures in this domain partially unfold as the barrier crossings occur.  
By comparison  the E domain remains more or less structurally intact even at the transition state which suggests that the relative immobility of 
this domain is crucial to the function of this biological namomachine \cite{ThirumARBBS01}. 
The dispersions in the TSE are also reflected in the heterogeneity of the distances between various salt-bridges in the transition states.  
The values of the contact distances, in the $T \rightarrow R$ transition  among the residues involved in  the salt-bridge switching 
between K80, R197, and E386 at the TS has a very broad distribution (Fig. 6-B)  
which also shows that the R197-E386 is at least 
partially disrupted in the TS and the K80-E386 is partially formed \cite{Horovitz02PNAS}.  \\

{\bf CONCLUSIONS}\\

Based on the observation that the conformational changes in molecular nanomachines are much slower than the rate of ligand binding-induced strain propagation, we have developed a structure-based method  to probe the allosteric transitions in biological molecules.  
In our method, transitions between multiple states are probed using the state-dependent energy functions and Brownian dynamics simulations.  
Applications to allosteric transitions in GroEL have produced a number of new predictions that can be experimentally tested.  The global dynamics in the $T \rightarrow R$ transition reveals that the A domain rotates in a counter-clockwise manner whereas they rotate in  a clockwise direction during the $R \rightarrow R^{\prime\prime}$ transition.  Although this observation was previously made based on the end point structures ($T$, $R$, and $R^{\prime\prime}$) alone, the transition kinetics show  substantial dynamic heterogeneity in the dynamics of individual subunits.  A key finding of the present work is that the transitions occur by a multiple coordinated switch between a network of salt-bridges.  Our findings suggest a series of mutational studies that can examine the link between disruption of the salt-bridge interactions and the GroEL function.  The most dramatic outside-in movement, the rearrangement of helices K and L of the A domain, occurs largely in the $R \rightarrow R^{\prime\prime}$ transition and  results in the  inter-subunit K80-D359 salt-bridge formation.  The TSEs  are broad with the $T \rightarrow R$ TSE  being more plastic than the one connecting $R$ and  $R^{\prime\prime}$ states. In both the transitions most of the conformational changes occur in the A domain with the E domain serving as a largely structurally unaltered static base that is needed for force transmission \cite{ThirumARBBS01}. 

The unprecedented picture of the dynamics of allostery in GroEL presented here  suggests that, like other ATP-consuming nanomachines, GroEL may be viewed as a force-generating device.  The combination of large-scale (nearly $\sim 50$ \AA)  dispersion in the SP binding sites  and the multivalent binding of  SP must axiomatically lead to 
considerable strain on the SP. 
Mechanochemical considerations enable us to make a rough estimate of the equilibrium force, $f$, 
acting on the SP resulting from the observed allosteric movements from 
$f\sim \frac{7\times (-\Delta G^{ATP}_h)}{\Delta d}$
where the ATP hydrolysis free energy $\Delta G^{ATP}_h(=\Delta G^o+RT\log{\frac{[ADP][P_i]}{[ATP]}})$ 
at physiological condition ($[ATP]=800 \mu M$, $[ADP]=80 \mu M$, $[P_i]=200 \mu M$, $T=320 K$, and $\Delta G^o=-7.3kcal/mol$) 
is $\sim -12 kcal/mol$, and $\Delta d \sim (10-50)$ \AA.  
Assuming an efficiency of about 0.5, we estimate $f\approx 5-30$ $pN$ which is large enough to partially or fully unfold many proteins.
\\        


{\bf METHODS}\\


\paragraph*{Energy function: }
Using the available structures in the Protein Data Bank (PDB) (PDB id: 1OEL ($T$ state) \cite{BrungerNSB95}, 2C7E ($R$ state) \cite{Ranson01Cell}, 1AON ($R^{\prime\prime}$ state) \cite{SiglerNature97}), we defined the \emph{state-dependent} SOP Hamiltonian \cite{HyeonSTRUCTURE06,HyeonBJ06_2} of the states ($X=T$, $R$, $R^{\prime\prime}$) of GroEL as  
\begin{eqnarray}
H(\{\vec{r}_i\}|X)&=&H_{FENE}+H_{nb}^{(att)}+H_{nb}^{(rep)}+H_{ATP}^{(att)}+H_{ATP}^{(rep)}+H_{sb}\nonumber\\
&=&-\sum_{i=1}^{N-1}\frac{k}{2}R_0^2\log({1-\frac{(r_{i,i+1}-r_{i,i+1}^o(X))^2}{R_0^2}})\nonumber\\
&+&\sum_{i=1}^{N-3}\sum_{j=i+3}^N\epsilon_h[(\frac{r^o_{ij}(X)}{r_{ij}})^{12}-2(\frac{r^o_{ij}(X)}{r_{ij}})^6]\Delta_{ij}(X)\nonumber\\
&+&\left[\sum_{i=1}^{N-2}\epsilon_l(\frac{\sigma}{r_{i,i+2}})^6+\sum_{i=1}^{N-3}\sum_{j=i+3}^N\epsilon_l(\frac{\sigma}{r_{ij}})^6(1-\Delta_{ij}(X))\right]\nonumber\\
&+&\sum_{i=1}^{N}\sum_{j\in ATP\ or\ ADP}\epsilon_h^{ATP}[(\frac{a^o_{ij}(X)}{r_{ij}})^{12}-2(\frac{a^o_{ij}(X)}{r_{ij}})^6]\Delta_{ij}(X)\nonumber\\
&+&\sum_{i=1}^N\sum_{j\in ATP\ or\ ADP}\epsilon_l^{ATP}(\frac{\sigma}{r_{ij}})^6(1-\Delta_{ij}(X))\nonumber\\
&+&\sum_{i^*,j^*\in\{\mathrm{salt\ bridge}\}}\frac{q_1q_2}{4\pi\epsilon r_{i^*j^*}}e^{-\kappa r_{i^*j^*}}.
\label{eq:SOP}
\end{eqnarray}
The first term, which accounts for chain connectivity is represented using  finite extensible nonlinear elastic (FENE) potential \cite{KremerJCP90}, with parameters
 $k=20$ $kcal/(mol\cdot$\AA$^2)$, $R_0=2$ \AA\ where
$r_{i,i+1}$ is the distance between neighboring interaction centers
$i$ and $i+1$, $r^o_{i,i+1}(X)$ is the
distance in state $X$.
The Lennard-Jones potential  (second term) accounts for interactions that stabilize
a particular allosteric state.
Native contact exists if the distance between $i$ and $j$
less than $R_C=8$ \AA\ in state $X$ for $|i-j|>2$.
If $i$ and $j$ sites are in contact in the native state, $\Delta_{ij}=1$, otherwise
$\Delta_{ij}=0$. To ensure the non-crossing of the chain, we used a $6^{th}$ power potential in the third and the fifth terms and set $\sigma=3.8$ \AA, 
which is the  $C_{\alpha}-C_{\alpha}$ distance.
We used $\epsilon_h=2$ $kcal/mol$ if the residues are in contact  and 
$\epsilon_l=1$ $kcal/mol$ for non-native pairs.

The fourth and the fifth terms in Eq.\ref{eq:SOP} are for  interaction of residues with ATP ($R$ state) or ADP ($R^{\prime\prime}$ state). 
The atomic coordinates of ATP (ADP) are taken from the $R$ ($R^{\prime \prime}$) structure without coarse-graining.    
The functional form for residue-ATP (or ADP) interaction is the same as for residue-residue interactions 
with $\epsilon_h^{ATP}=0.2kcal/mol$, $\epsilon_l^{ATP}=0.1kcal/mol$. 
We used a small value of $\epsilon_h^{ATP}$ (or $\epsilon_l^{ATP}$)  because  the coordinates of all the
heavy atoms of ATP and ADP  are explicitly used  as interaction sites.  
The distance between the $i^{th}$ residue and the $j^{th}$ atom in ATP (or ADP) is $a_{ij}$. 
We used
the screened electrostatic potential where $\kappa^{-1}=2.4$ \AA, $\epsilon=10\epsilon_0$, 
and $q_1q_2=-e^2$ to account for the favorable salt-bridge interactions which are state-independent (Eq.\ref{eq:SOP}).  


\paragraph*{Inducing allosteric transitions:}
The $T\rightarrow R$ allosteric transition of GroEL is simulated by integrating the equations of motion with the force arising from  $H(\{\vec{r}_i\}|R)$. However, the ensemble of initial
structures were generated using the T-state $H(\{\vec{r}_i\}|T)$.    The Brownian dynamics algorithm \cite{McCammonJCP78,VeitshansFoldDes96} determines the configuration of GroEL at time $t$ as follows, 
\begin{equation}
\begin{array}{lll}
     (i)& \vec{r}_i(t+h)=\vec{r}_i(t)+\frac{h}{\zeta}(\vec{F}_i(t|T)+\vec{\Gamma}_i(t))& \mbox{$(0\leq t< t^*)$}\\
     (ii)& \vec{r}_i(t+h^*)=\vec{r}_i(t)+\frac{h^*}{\zeta}(\vec{F}_i(t|T\rightarrow R)+\vec{\Gamma}_i(t))& \mbox{$(t^*\leq t <t^*+N_th^*)$}\\
     (iii)& \vec{r}_i(t+h)=\vec{r}_i(t)+\frac{h}{\zeta}(\vec{F}_i(t|R)+\vec{\Gamma}_i(t))& \mbox{$(t\geq t^*+N_th^*)$}
\end{array}
\label{eqn:Brownian}
\end{equation}
where $\vec{F}_i(t|X)=-\nabla_{\vec{r}_i}H(\{\vec{r}_i\}|X)$ ($X=T$, $R$ or $T\rightarrow R$), the Newtonian force acting on a residue $i$,
and $\Gamma_i(t)$ is a random force on $i^{th}$ residue that has a white noise
spectrum that satisfies  
$\langle\vec{\Gamma}_i(t)\cdot\vec{\Gamma}_j(t+nh)\rangle=\frac{6\zeta k_BT}{h}\delta_{0,n}\delta_{i,j}$, 
where $\delta_{0,n}$ is the Kronecker delta function and $n=0,1,2,\ldots$...  As long as the fluctuation-dissipation theorem is satisfied it can be shown that our
 procedure for switching  the Hamiltonian from $T$ to $R$ will lead to the correct Boltzmann distribution for the $R$ state at long times \cite{GardinerBook}. 

The algorithm in Eq.\ref{eqn:Brownian} is implemented in three steps: 
(i) During the time interval  $0\leq t<t^*$ an ensemble of T-state conformations is generated;  
(ii) The energy function is switched from $H(\{\vec{r}_i\}|T)$ to 
$H(\{\vec{r}_i\}|R)$ symbolized by $H(\{\vec{r}_i\}|T\rightarrow R)$ in the duration
$t^*\leq t <t^*+N_th^*$. If $N_t = 0$ our method is similar to one in \cite{Koga06PNAS}; 
(iii) A dynamic trajectory 
under $H(\{\vec{r}_i\}|R)$ is generated for $t >t^*$.
The assumption in our method is that the rate of conformational change in biomolecules is smaller than the rate at which a locally applied strain (due to ligand binding) propagates. As a result,   the Hamiltonian switch should not be instantaneous ($N_t\neq 0$).  Using a non-zero value of $N_t$ (second step in Eq.\ref{eqn:Brownian}) not only ensures that 
there is a lag time between ligand binding
and the associated response but also eliminates computational instabilities in the distances between certain residues that change dramatically during the transition.  The "loading" rate can be altered by varying $N_t$ and hence even non-equilibrium ligand-induced transitions can be simulated.  Additional details of the simulations are given in the Supplementary Information.


\paragraph*{Additional details: }
When the transition, say $T$ to $R$ occurs,  
the native contact distances ($r^o_{ij}(T)$ and $r^o_{ij}(R)$) in the two states are different. 
If $r^o_{ij}(T)\ll r^o_{ij}(R)$ then, upon the Hamiltonian switch,  a large repulsive force arises from the the pair $i$ and $j$ that is initially in the equilibrated $T$-state conformation at a distance, $r_{ij}$ $(\approx r^o_{ij}(T))$.
To eliminate the purely computational problem 
we made a gradual transition of the parameters using a linear interpolation procedure.  
For the $T\rightarrow R$ transition we used,
$r_{ij}^o(T\rightarrow R)=\frac{(K-k)r_{ij}^o(T)+kr_{ij}^o(R)}{K}$. 
The $a_{ij}^o(T\rightarrow R)$values are similarly defined with  $a_{ij}^o(T)=0$ but $a_{ij}^o(R\ or\ R'')\neq 0$. 
Accordingly, the corresponding Hamiltonian 
$H(\{\vec{r}_i\}|T\rightarrow R)$ for stage (ii) is defined by inserting 
$r_{ij}^o(T\rightarrow R)$, $a_{ij}^o(T\rightarrow R)$ in Eq.\ref{eq:SOP}. 
We set $K=100$ and increased $k$ from 0 to $K$ every $\delta=100$ integration time step, which leads to $N_t=K\times\delta$. 
In addition, we choose the integration time step $h^*=(0.001-0.01)\tau_L$ during the Hamiltonian switch. 
Thus, the Hamiltonian switch is smoothly made within $(5-50)$ $ns$ without causing any computational instability.
The technical
problem arises only for certain pairs of residues for which $r^o_{ij}(T)\ll r^o_{ij}(R)$.\\


{\bf ACKNOWLEDGEMENTS} This work was supported in part by a grant from the National Institutes of Health(1R01GM067851-01).\\

\newpage

\newpage 
{\bf FIGURE CAPTIONS}\\

{{\bf Figure 1} :}
GroEL structure. 
The columns from left to right show $T$, $R$, and $R^{\prime\prime}$ structures of GroEL structures.  
The  top view is given in first row (see Fig. S1 in Supplementary Information for a side view) and  the second row displays the side view of a single subunit. The white ball represents D359. 
The helices that most directly influence the allosteric transitions transitions are labeled. 

{{\bf Figure 2} :}
GroEL dynamics monitored using various angles. 
{\bf A}. $T\rightarrow R$ transition dynamics for the heptamer monitored using angles, $\alpha$, $\beta$, $\gamma$. 
An angle $\theta$ ($=\alpha$, $\beta$) 
is defined by $\cos{\theta(t)}=\vec{u}_{\theta}(0)\cdot\vec{u}_{\theta}(t)/|\vec{u}_{\theta}(0)||\vec{u}_{\theta}(t)|$. 
For $\alpha$, we obtain $\vec{u}_{\alpha}(t)$ by projecting the vector ($\vec{r}_{236(i)}(t)=\vec{R}_{236(i)}(t)-\vec{R}_{CM}$) between the center of mass ($\vec{R}_{CM}$) and  residue 236 on $i^{th}$ subunit ($\vec{R}_{236(i)}(t))$ onto the plane perpendicular to the principal axis ($\hat{e}_P$) of the heptamer, i.e.,. 
$\vec{u}_{\alpha}(t)=\vec{r}_{236(i)}(t)-(\vec{r}_{236(i)}(t)\cdot\vec{e}_P)\vec{e}_P$. 
The angle between H helices (residue 231$-$242) of $i^{th}$ subunit at times $t=0$ and $t$ using the vector, $\vec{R}_{231(i)}(t)-\vec{R}_{242(i)}(t)$ is $\beta$. The sign of the angles ($\alpha$ and $\beta$) is determined using 
$sgn[(\vec{u}(0)\times\vec{u}(t))\cdot \hat{e}_P]$, which is (+) for counter-clockwise and (-) for clockwise rotation.  
$\gamma$ measures the perpendicular motion of apical domain with respect to the hinge (residue 377). We defined $\vec{u}_{\gamma}(t)=\vec{R}_{236(i)}(t)-\vec{R}_{377(i)}(t)$ at each subunit $i$, and $\gamma(t)=90^o-\cos^{-1}{(\vec{u}_{\gamma}\cdot\hat{e}_P)}$.
On the right three panels we plot the time dependence of $\alpha$, $\beta$, and $\gamma$ for each subunit in different color. 
The black line represents the average of  21 ($=3\times 7$) values of each angle calculated from three trajectories of 7 subunits.  
{\bf B}. Same as in {\bf A} except for the $R\rightarrow R^{\prime\prime}$ transition.

{{\bf Figure 3} :}
RMSD as a function of time. 
{\bf A.} Time-dependence of RMSD  of a few individual molecules are shown for $T\rightarrow R$ transition. 
Solid (dashed) lines are for RMSD/$T$ (RMSD/$R$)  (RMSD calculated with respect to the $T$ ($R$) state). 
The enlarged inset gives an example of a trajectory, in blue, that exhibits multiple passages across the transition region. 
{\bf B.} Ensemble averages of the RMSD for the $T\rightarrow R$ (top) and $R\rightarrow R^{\prime\prime}$ (bottom) transitions are obtained over 50 trajectories. 
The solid lines are exponential fits to $RMSD/R$ and $RMSD/R^{\prime\prime}$  relaxation kinetics. {\bf C.} Time-dependent changes in the angles (measured with respect to the $T$ state) that  F, M helices make during the $T\rightarrow R\rightarrow R^{\prime\prime}$ transitions. The inset shows the dispersion of individual trajectories for F-helix  with the black  line being the average.
{\bf D.} Time-dependent changes in the angles (measured with respect to the $T$ state) that  K, L helices make during the $T\rightarrow R\rightarrow R^{\prime\prime}$ transitions. The inset on the top shows the structural changes in  K, L helices during the $T\rightarrow R\rightarrow R^{\prime\prime}$ transitions. For clarity, residues 357-360 are displayed in space-filling representation in white.  The bottom inset shows the dispersion of individual trajectories for the K-helix.  The black  line is the average. In {\bf C} and {\bf D} $\theta = cos^{-1}(\vec{u}(0)\cdot\vec{u}(t))$.


{{\bf Figure 4} :}
$T\rightarrow R$ GroEL dynamics monitored using of two interacting subunits. 
Side views from outside to the center of the GroEL ring and top views are presented for the $T$ (left panel) and $R$ (right panel) states. 
Few residue pairs  are annotated and connected with dotted lines.  
The ensemble average kinetics of a number of salt-bridges and contacts between few other residues are shown in the middle panel.  Distance changes for a single trajectory for few residues are given in Fig. S4 in the Supplementary Information. Fits of the relaxation kinetics are: 
$\langle d(t)\rangle_{R58-E209}/\mathrm{\AA}=14.9+9.6(1-0.17e^{-t/5.1\mu s}-0.83e^{-t/825\mu s})$,
$\langle d(t)\rangle_{D83-K327}/\mathrm{\AA}=8.5+4.9(1-e^{-t/100.0\mu s})$,
$\langle d(t)\rangle_{P33-N153}/\mathrm{\AA}=7.3+4.2e^{-t/6.3\mu s}$, 
$\langle d(t)\rangle_{R284-E386}/\mathrm{\AA}=13.2+16.5(1-0.49e^{-t/20.8\mu s}-0.51e^{-t/85.8\mu s})$, 
$\langle d(t)\rangle_{R285-E386}/\mathrm{\AA}=12.6+15.8(1-0.42e^{-t/19.1\mu s}-0.51e^{-t/88.8\mu s})$, 
$\langle d(t)\rangle_{R197-E386}/\mathrm{\AA}=11.9+9.0(1-0.29e^{-t/0.67\mu s}-0.71e^{-t/96.7\mu s})$, 
$\langle d(t)\rangle_{K80-E386}/\mathrm{\AA}=10.4+9.8(0.78e^{-t/12.1\mu s}+0.22e^{-t/61.8\mu s})$,
$\langle d(t)\rangle_{E257-R268}/\mathrm{\AA}=9.7+12.1(1-0.35e^{-t/26.2\mu s}-0.65e^{-t/66.4\mu s})$. 
\noindent Initially, the dynamics of salt-bridge formation between E257 and K321, R322, K245 show non-monotonic behavior. 
Thus, we did not perform  a detailed kinetic analysis for these residues. 

{{\bf Figure 5} :}
Dynamics of the $R\rightarrow R^{\prime\prime}$ transition using two-subunit SOP model simulations. The dynamics along one trajectory are shown in Fig. S4 in the Supplementary Information.  Intra-subunit salt-bridges (or residue pairs) of interest (D83-K327, E409-R501, P33-N153) are plotted on the top panel, and inter-subunit salt-bridges (or residue pairs) of interest (E257-K246, E257-R268, E257-K321, E257-R322, I305-A260) are plotted on the bottom panel. For emphasis, K80-D359 salt-bridge dynamics, that provides a driving force to other residue dynamics, is specially plotted on the bottom panel.
The quantitative kinetic analysis performed for rupture of D83-K327 and formation of K80-E359 salt-bridges show 
$\langle d(t)\rangle_{D83-K327}/\mathrm{\AA}=10.4+26.9(1-e^{-t/77.9\mu s})$, 
$\langle d(t)\rangle_{K80-D359}/\mathrm{\AA}=14.1+26.4e^{-t/28.0\mu s}$.

{{\bf Figure 6} :}
Transition state ensembles (TSE). 
{\bf A}. TSEs are represented in terms of distributions $P(q^{\ddagger})$ where  
$q^{\ddagger}\equiv\frac{\Delta^{\ddagger}-min(RMSD/X)}{max(RMSD/X)-min(RMSD/X)}$.  
Histogram in red gives $P(q^{\ddagger})$  for $T\rightarrow R$ (red) and the data in green are 
for the $R\rightarrow R^{\prime\prime}$ transitions.   
For $T\rightarrow R$, $X=R$, $min(RMSD/X)=1.5$ \AA\ and $max(RMSD/X)=8.0$ \AA. 
For $R\rightarrow R''$, $X=R^{\prime\prime}$, $min(RMSD/X)=1.5$ \AA\ and $max(RMSD/X)=14.0$ \AA. 
To satisfy conservation of the number of molecules  
the distributions are normalized using
$\int dq^{\ddagger}\left[P(q^{\ddagger}|T\rightarrow R)+P(q^{\ddagger}|R\rightarrow R")\right]=1$. 
 Twenty overlapped TSE structures  for the two transitions are displayed.   
In the bottom panel, the distributions of $t_{TS}$ that satisfy  $\delta^{\ddagger} < 0.2$ \AA, 
are plotted for the $T\rightarrow R$ and the $R\rightarrow R''$ transitions. 
{\bf B}. For the $T\rightarrow R$  TSE we show   
the salt-bridge distances $(d^{R197-E386}_{TS},d^{K80-E386}_{TS})$ with black dots. 
The red and the green dots are the equilibrium distances 
$(\langle d^{R197-E386}_{TS}\rangle,\langle d^{K80-E386}_{TS}\rangle)$ in the $T$ and the $R$ states, respectively. 
The distance distributions for the TSE are shown in blue.


\clearpage 
\begin{figure}[ht]
\includegraphics[width=6.00in]{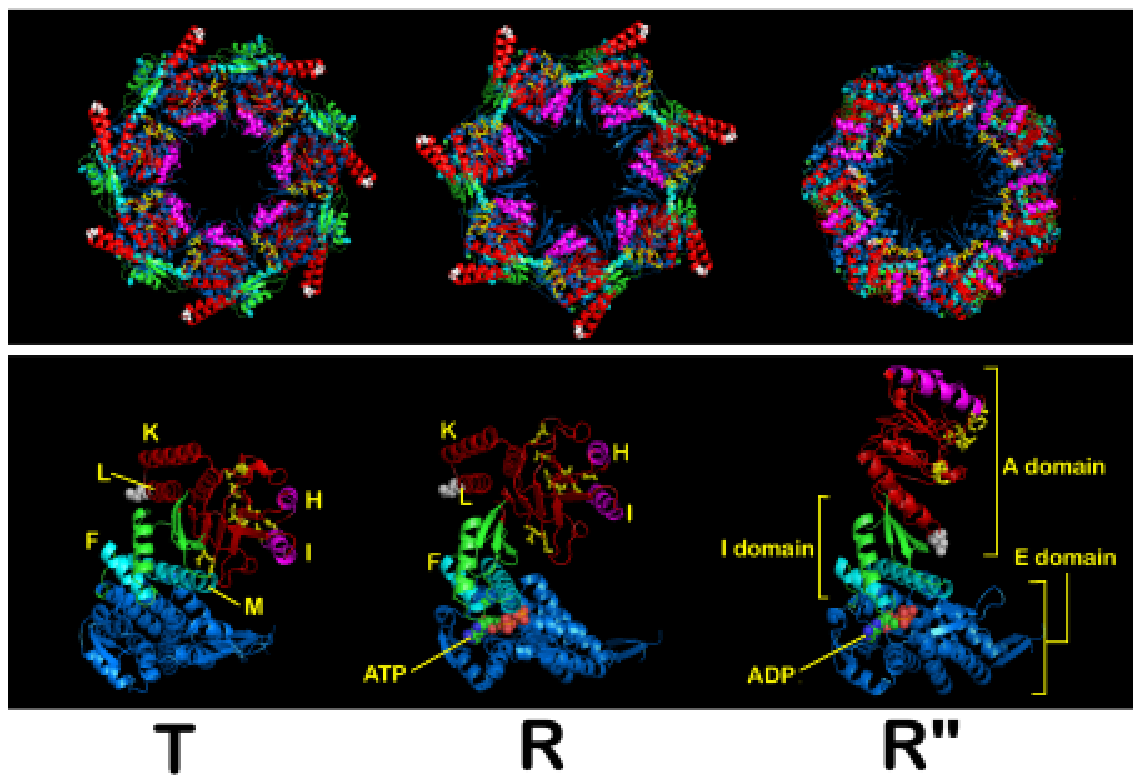}
\caption{\label{GroEL_structure.fig}}
\end{figure}

\begin{turnpage}
\begin{figure}[ht]
\includegraphics[width=9.00in]{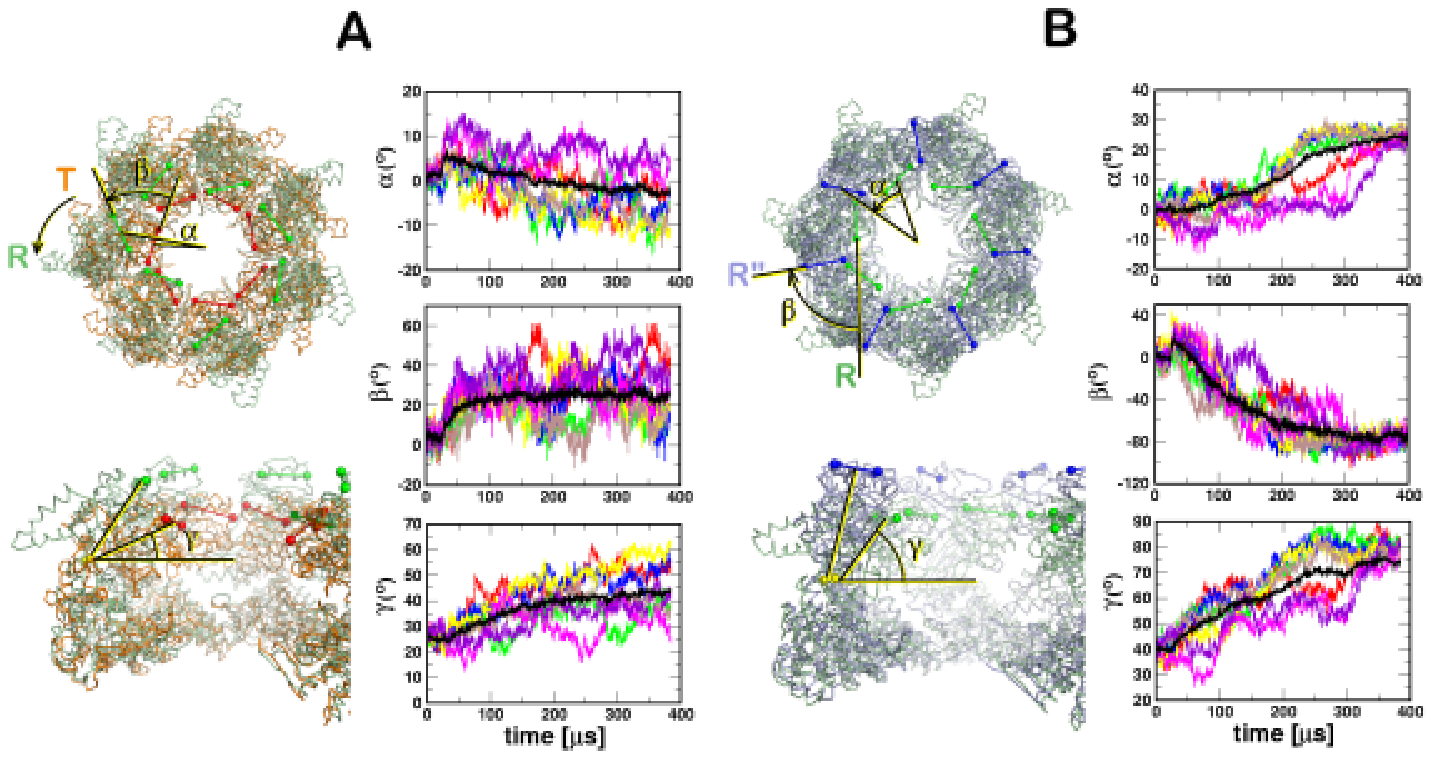}
\caption{\label{TRRpp_angles.fig}}
\end{figure}
\end{turnpage}

\begin{figure}[ht]
\includegraphics[width=7.00in]{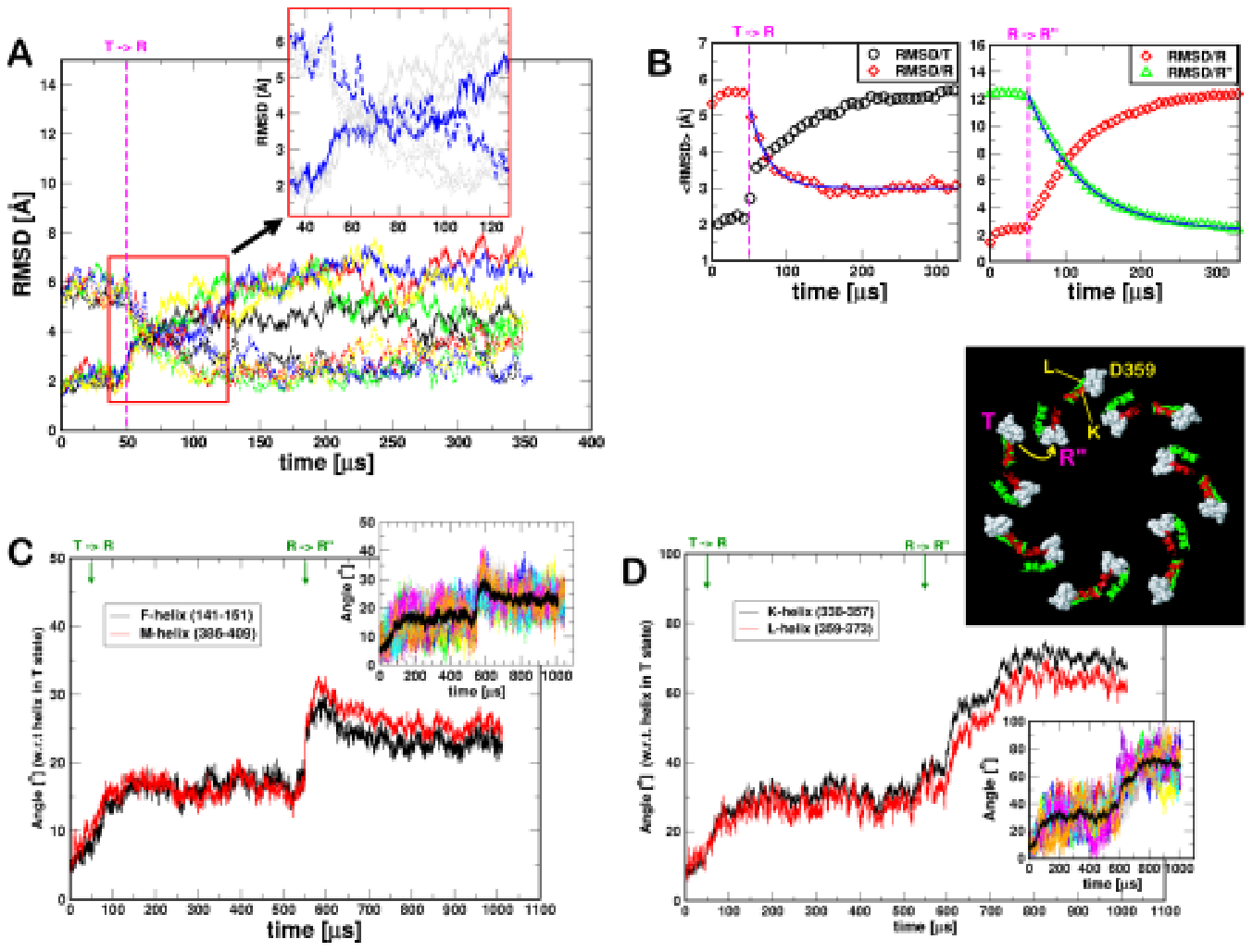}
\caption{\label{RMSD_and_angle.fig}}
\end{figure}

\begin{turnpage}
\begin{figure}[ht]
\centering
\includegraphics[width=8.00in]{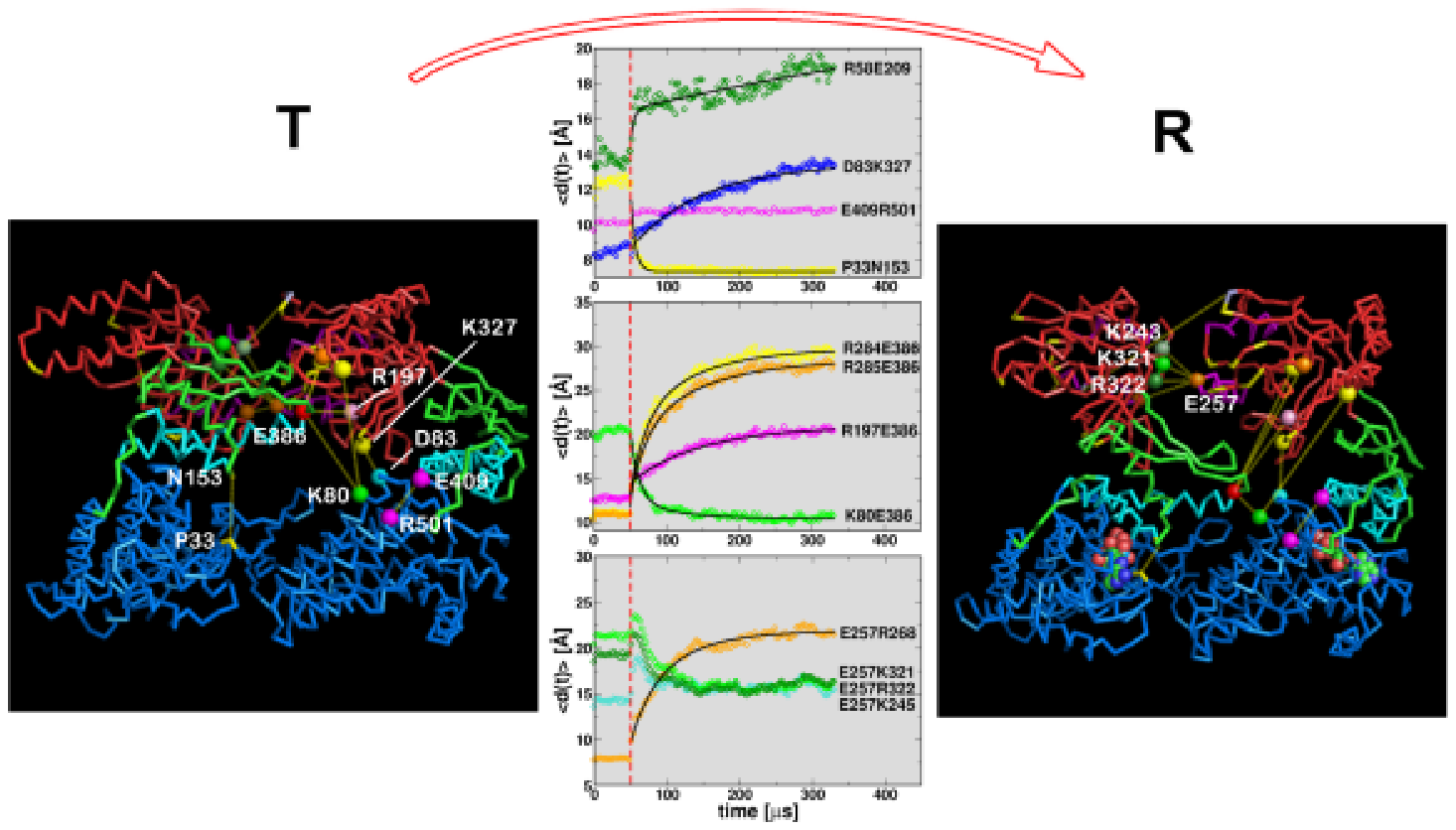}
\caption{\label{saltbridgeanalysisTR.fig}}
\end{figure}
\end{turnpage}

\begin{turnpage}
\begin{figure}[ht]
\centering
\includegraphics[width=8.00in]{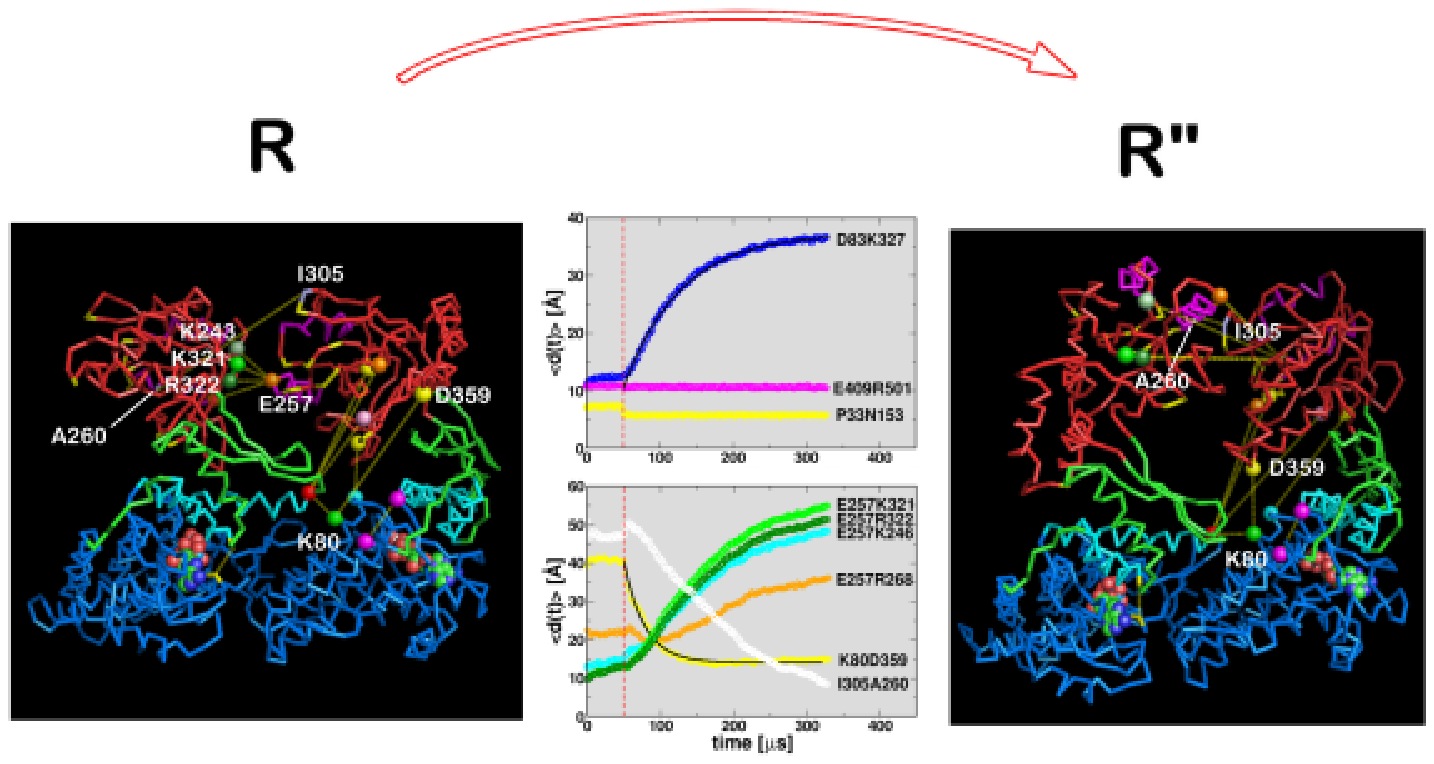}
\caption{\label{saltbridgeanalysisRRpp.fig}}
\end{figure}
\end{turnpage}

\begin{turnpage}
\begin{figure}[ht]
\includegraphics[width=6.00in]{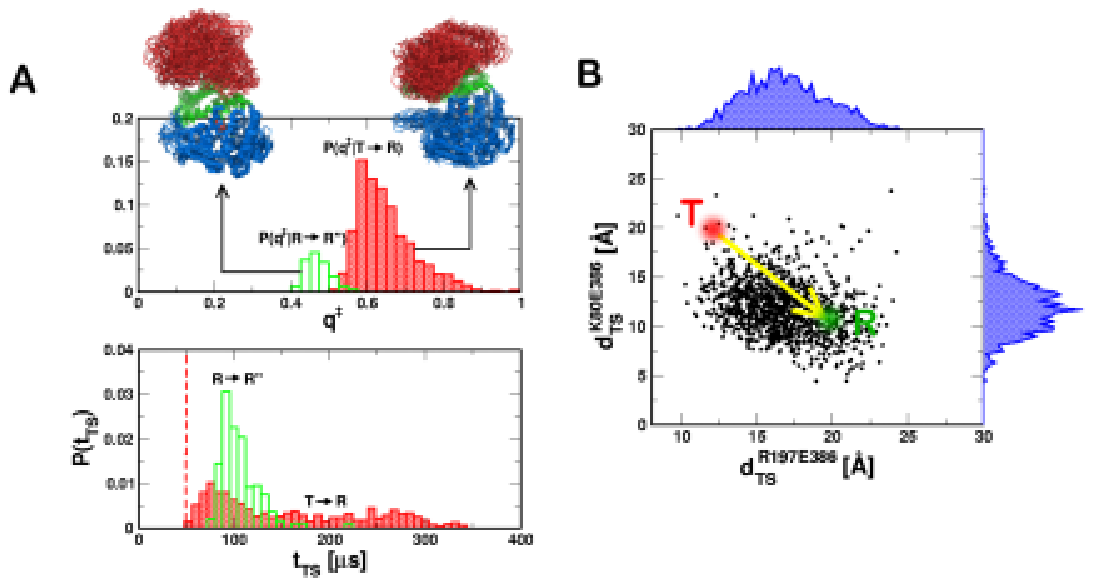} 
\caption{\label{TSE.fig}}
\end{figure}
\end{turnpage}
\clearpage


\section*{SUPPLEMENTARY INFORMATION}

{\bf Rationale for using the Self-organized Polymer (SOP) model for allosteric transitions:} Several studies have shown that the gross features of protein folding \cite{OnuchicCOSB04,HyeonBC05}, 
mechanical unfolding of proteins and RNA \cite{Klimov2,HyeonPNAS05} and the complicated global motions of large systems can be captured 
using native structure-based models \cite{Bahar05COSB}.  
More recently, such native structure-based methods \cite{Miyashita03PNAS,Maragakis05JMB,Best05Structure,Koga06PNAS,Okazaki06PNAS} have also been used to probe transitions between two given end-point structures and to ascertain the nature of robust modes that mediate the allosteric transitions \cite{Zheng05Structure}.   
In order to realistically simulate transitions between distinct allosteric states,
it is necessary to use simple coarse-grained models that can be used to capture, at least qualitatively, 
the inherent dynamics connecting two or more states. 
Although simple elastic network models have given insights into the nature of low frequency dynamics of a number of systems \cite{Bahar05COSB}, the inherent linearity of the model limits their scope when dealing with potential non-linear motions that must be involved in the allosteric transitions \cite{Miyashita03PNAS}.  We have recently introduced a novel class of versatile structure-based models that incorporates the fundamental polymeric aspects of proteins and RNA. The SOP model, which is easily adopted for use in very large systems, has already proven to be successful in obtaining a number of totally unanticipated results for forced-unfolding and force-quench refolding of RNA and proteins \cite{HyeonSTRUCTURE06,HyeonBJ06_2}.  

Building on the successful application to single molecule force spectroscopy of biomolecules we used a variant of the SOP model to probe the complex allosteric transitions in a prototypical biological nanomachine, namely, the \textit{E. Coli.} chaperonin GroEL.   
Structures of GroEL ($T$, $R$, and $R^{\prime\prime}$), that are populated in the reaction cycle, have been determined (see Fig. S1 for a side view). Given the two allosteric states (say $T$ and $R$ of GroEL) we induce transitions between the two states by switching the energy function representing one structure to another.  A few methods for achieving the switch smoothly have been recently proposed \cite{Maragakis05JMB,Best05Structure,Koga06PNAS,Okazaki06PNAS}.  
We have advocated a Langevin dynamics based method in which the switch is accomplished using Eq. 2 of the main text (see Fig. S2 for a pictorial view).  In writing the equations of motions given by Eq. 2 
in the main text, we assume that initially the ensemble of conformations are in the $T$ state so that the conformations obey the Boltzmann distribution i.e, $P(\{\vec{r}_i\}) \simeq e^{-\beta H(\{\vec{r}_i\}|T)}$ 
where $H(\{\vec{r}_i\}|T)$ is the SOP energy  for the $T$-state GroEL (see Eq. 2 in the main text).  
Upon switching the Hamiltonian (see below) over a time interval $N_th^*$ the dynamics follows the equations of motion given in step (iii) of 
Eq. 2 in the main text.  
As long as fluctuation dissipation theorem is satisfied, so that the standard potential conditions are met \cite{GardinerBook},  
it can be rigorously shown that at long times the biomolecule will reach the $R$ state with the conformations given by the equilibrium Boltzmann distribution governed by the Hamiltonian  $H(\{\vec{r}_i\}|R)$. 
From this perspective the procedure we have used is rigorous.

The assumption that switch occurs over a predetermined duration requires a few comments.  The allosteric transitions occur as a result of ligand-binding or interactions with other biomolecules.  As a result of binding, local strain is induced at the interaction site or sites.  As long as the rate of strain propagation is larger than the rate of conformational change of the molecule then the switch over a reasonable time is justified.  Extremely rapid switching, which is tantamount to very large local loading rates, is unphysical as is adiabatic change in the energy function.  Given these extreme situations we choose a value of $N_t$ that falls in between the extreme conditions.  In our procedure $N_t$ can be adjusted to mimic the potential rate of strain propagation, which induces the allosteric transitions.  The efficacy of the procedure has been demonstrated by successful applications to describe the multiple allosteric  transitions in GroEL.
\\

{\bf Implementation of the Hamiltonian switch:}  Here we give details of the algorithm for executing the second step in the equations of motion (See subsection \textit{Additional details} in the Methods section in the main text).  During the 
 transition interval we define $r^o_{ij}(T\rightarrow R)$ using  linear
combination of 
$r^o_{ij}(T)$ and $r^o_{ij}(R)$ where $r^o_{ij}(X)$ is the distance between the residues $i$ and $j$ in the structure $X$ with $X = T$ or $R$. 
The switch between $r^o_{ij}(T)$ and $r^o_{ij}(R)$ is carried out slowly  every
100 time steps. 
In the initial stage of  the $T\rightarrow R$ transition (0 to 100 time steps) we let 
$r^o_{ij}(T\rightarrow R)=r^o_{ij}(T)$. Subsequently, we set  $r^o_{ij}(T\rightarrow R) = (1 -0.01k)r^o_{ij}(T) + 0.01 k r^o_{ij}(R)$ where $k = 1, 2, 3....100$ is changed every 100 time steps.  Thus, the switch in $r^o_{ij}(T \rightarrow R)$ occurs over 10,000 time steps.  The loading condition can be varied by changing the number of time steps used to achieve the switch in the distances between the native contacts.  For convenience we used a linear combination of $r_{ij}^o(T)$ and $r_{ij}^o(R)$ during the switch process. More generally, one can use non linear combinations, i.e., $r_{ij}^o(T \rightarrow R) = g(t)r_{ij}^o(T) + (1 - g(t))r_{ij}^o(R)$ where $g(t)$ is an arbitrary function (exponential for example).

At the end of each interval, the new value of $r^o_{ij}(T\rightarrow R)$  is
substituted into the SOP Hamiltonian to compute the forces needed to solve the equations of motion in step (ii) (see Eq. 2 of the main text). 
In the present application, the procedure for using $r^o_{ij}(T \rightarrow R)$ lasts only $<50$ $ns$. 
As a result, the dynamics of distant pair is not affected. 
Only the equilibrium distances of native pairs, that are already in
contact in the $T$ state but lead to instability in the intergration of equations of motion due to rapid switching
from $T$ to $R$, are corrected to the equilibrium distances at $R$ state (see Fig. S2).  \\

{\bf Time scales and their relevance:}  The characteristic time scale of the Brownian dynamics in the  overdamped limit is  
$\tau_H=\frac{\zeta\epsilon_h h}{k_BT}\tau_L$ 
where we used the friction coefficient $\zeta=50\tau_L^{-1}$, $\epsilon_h=2.0 kcal/mol$ and $\tau_L=(\frac{ma^2}{\epsilon_h})^{1/2}\sim 3$ $ps$ for proteins.
 The simulations are performed at $T=300$ $K$ ($k_BT=0.6kcal/mol$). We chose the integration time step $h=0.1\tau_L$ for (i) and (iii) while 
$h^*=(0.001-0.01)\tau_L$ for (ii).
Thus, $10^6$ integration time steps with $h=0.1\tau_L$ in our Brownian dynamics simulations correspond to $50$ $\mu s$.  Because we have used a minimal model for GroEL the time scales quoted in the main text should be viewed as lower limit for the various processes. The actual time scales are expected to be much longer. The relatively time scales for different aspects of the allosteric transitions are likely to be correct. For example, we find that the tilt of K and L helices occur four times more slowly than the F and M helices (see Fig. 3 in the main text).  Our prediction of factor of four is, in all likelihood, an accurate estimate.  
During simulations we collected the structures every 0.5 $\mu s$ to analyze the allosteric transition dynamics. \\

\paragraph*{Analysis of dynamics:}
To perform a quantitative analysis on the salt bridge or contact pair dynamics 
we averaged over the time traces of all the trajectories. 
For the contact dynamics of two-subunit GroEL we generated $N=50$ trajectories in total and computed dynamic changes in specific residue pairs
using
$\langle d(t)\rangle=\frac{1}{N}\sum_{i=1}^Nd_i(t)$. In general  $\langle d(t)\rangle$ is fit using  
$\langle d(t)\rangle=\langle d(t^*)\rangle+\Delta(fe^{-(t-t^*)/\tau_1}+(1-f)e^{-(t-t^*)/\tau_2})$, where $t^*=50\mu s$, $\Delta$ is the average decrease in  the 
contact distance, and $\tau_1$, $\tau_2$ are the relaxation times for the pathways partitioned into $f$ and $1-f$. \\


\clearpage
\noindent {\bf Figure Captions} \\

{\bf Figure S1:} 
The columns from left to right show the side view of GroEL structure in the $T$, $R$, and $R^{\prime\prime}$ states.

{\bf Figure S2: }
Illustration of the procedure to switch SOP Hamiltonian from $T$ to $R$ state. 
To avoid the computational instability caused by instantaneous switch of equilibrium distance, the equilibrium distance from $T$ to $R$ state ($r^o_{ij}(T)\rightarrow r^o_{ij}(R)$) is gradually switched using a series of transient potentials defined with $r^o_{ij}(T\rightarrow R)$ (see \emph{Additional details} in Methods section). 

{\bf Figure S3: }
TSEs represented in terms of distribution $P(\Delta^{\ddagger})$ where 
$\Delta^{\ddagger}=1/2\times|(RMSD/T)(t_{TS})+(RMSD/R)(t_{TS})|$ for $T\rightarrow R$ transition. $\Delta^{\ddagger}$ for $R\rightarrow R^{\prime\prime}$ transition is similarly defined. 

{\bf Figure S4: }
The dynamical changes in the distances between a number of residues in a single trajectory during $T\rightarrow R$ and $R\rightarrow R^{\prime\prime}$ are plotted on {\bf A} and {\bf B}, respectively. 

\clearpage
\renewcommand{\thefigure}{S1}
\begin{figure}
\centering
\includegraphics[width=5.0in]{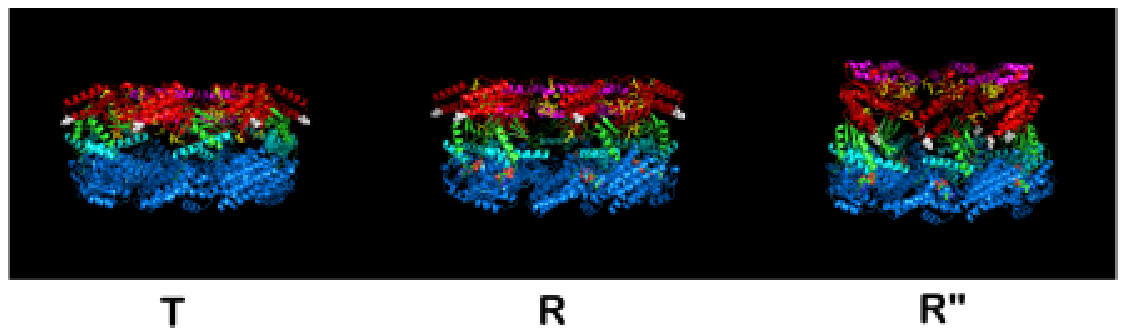}
\caption{}
\end{figure}
\clearpage
\renewcommand{\thefigure}{S2}
\begin{figure}
\centering
\includegraphics[width=5.0in]{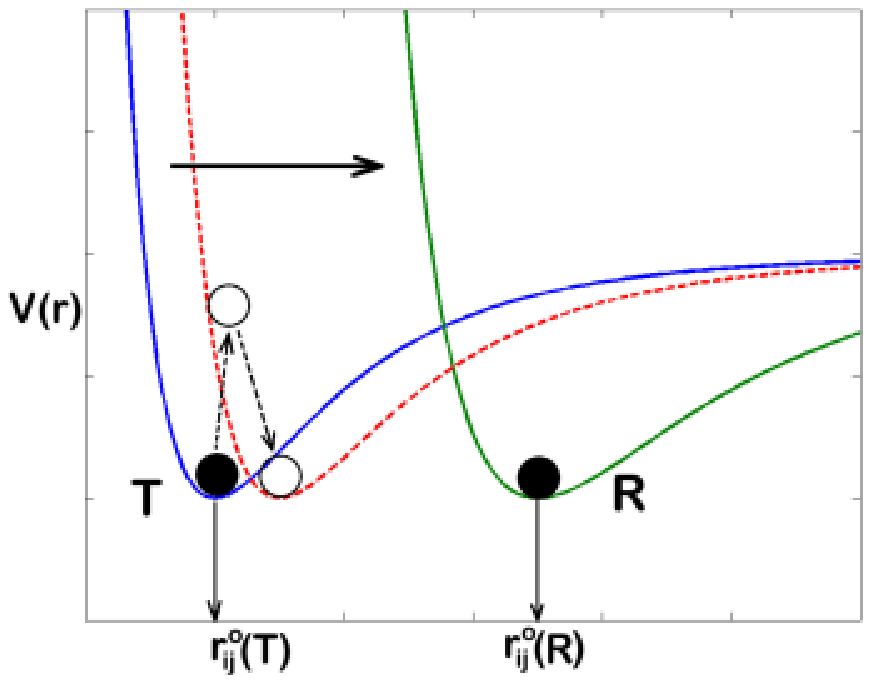}
\caption{}
\end{figure}
\clearpage
\renewcommand{\thefigure}{S3}
\begin{figure}
\centering
\includegraphics[width=5.0in]{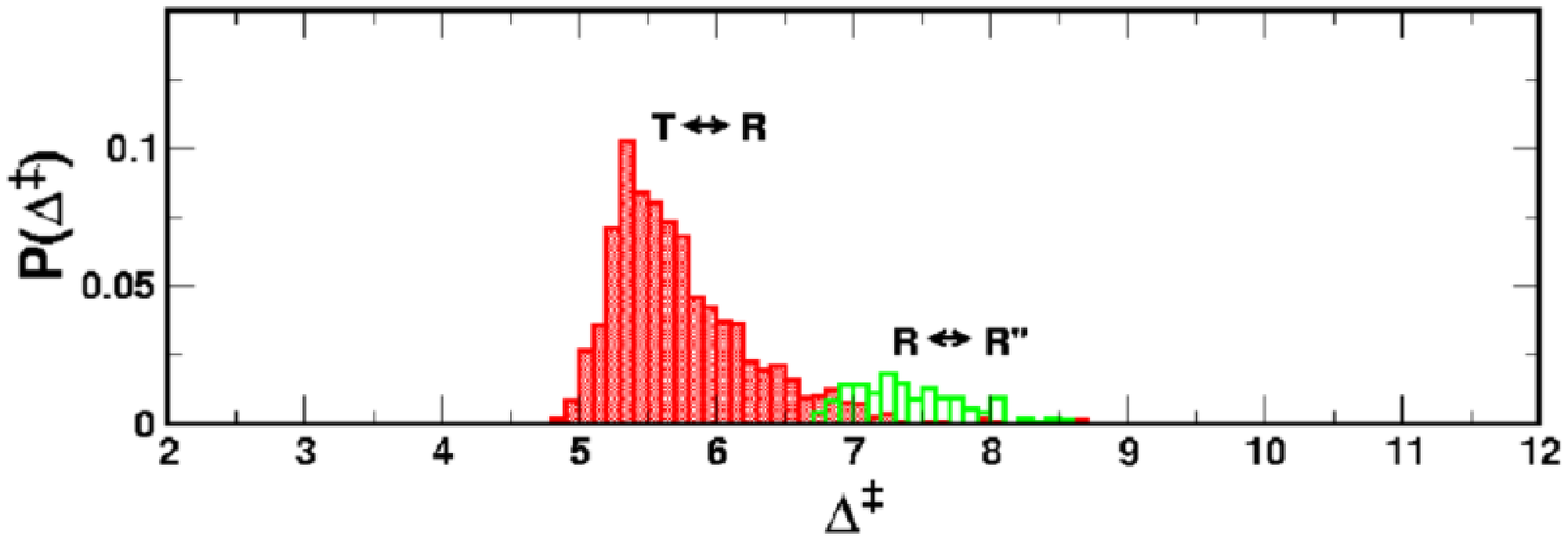}
\caption{}
\end{figure}
\renewcommand{\thefigure}{S4}
\begin{figure}
\centering
\includegraphics[width=4.5in]{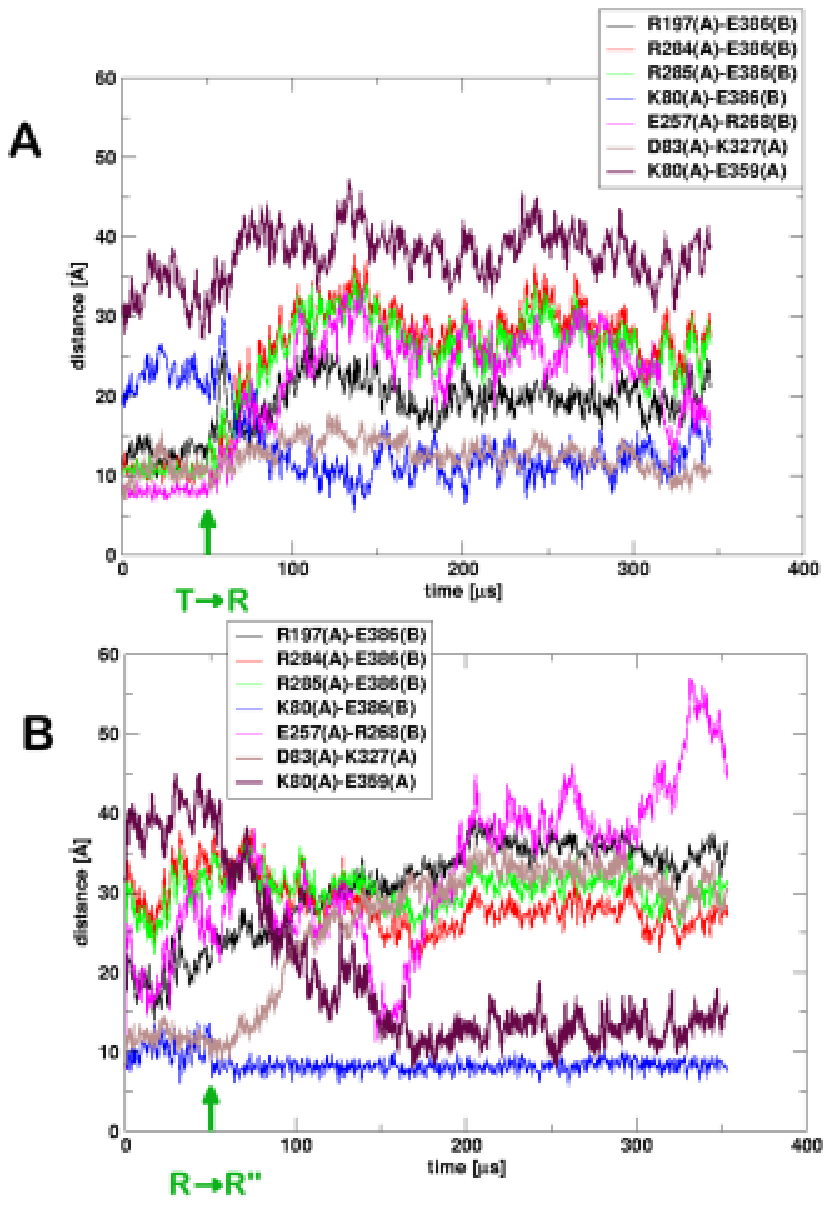}
\caption{}
\end{figure}

\end{document}